# ZERO AND FIRST ORDER LAMB AND SH WAVES PROPAGATION IN LANGASITE SINGLE CRYSTAL PLATES UNDER THE INFLUENCE OF DC ELECTRIC FIELD

S.I. Burkov[1], O.P. Zolotova[1], B.P. Sorokin[2], P.P. Turchin[1]

1) Siberian Federal University, 79 Svobodny ave., 660041 Krasnoyarsk, Russia
2) Technological Institute for Superhard and Novel Carbon Materials, 7a, Centralnaya str., 142190 Troitsk, Moscow region, Russia

e-mail: sergbsi@gmail.com; bpsorokin2@rambler.ru

*Abstract*— Paper is presented the results of computer simulation. Effect of the homogeneous dc electric field influence on the propagation of zero and first order Lamb and SH waves in piezoelectric langasite single crystal plates for a lot of cuts and directions have been calculated. Crystalline directions and cuts with maximal and minimal influence of dc electric field have indicated. Effect of hybridization of plate modes has been discussed.

In commemoration of academician of RAS K.S. Aleksandrov

## INTRODUCTION

Langasite single crystals (LGS, $La_3Ga_5SiO_{14}$) and its isomorphs belonging to the 32 point symmetry group are of great importance because the successful combination of a number of useful properties such as thermostability, small attenuation of acoustic waves and a better electromechanical coupling in comparison with quartz [1, 2]. LGS thermostable cuts which can be used to produce the shear wave resonators have the frequency coefficient which is 20 % smaller than that of AT-cut quartz resonators. This fact combining with electromechanical coupling coefficient (EMCC) greater than that of AT-cut allows the considerable decreasing (up to 1.5-2.0 times) of the linear dimensions of a piezoelectric resonator. Additionally langasite crystals possess a good chemical stability and a lack of any phase transitions up to melt point (1470 °C). Last circumstance gives a possibility of high temperature devices fabrication [3, 4].

Earlier authors [5, 6] have proposed a methods of experimental observation of homogeneous dc electric field or mechanical pressure on bulk acoustic wave (BAW) propagation and as a consequence the measuring of the complete set of nonlinear electromechanical properties for such piezoelectric crystals as bismuth germanium oxide, bismuth silicium oxide and langasite. Calculation of above mentioned influences on the BAW and SAW propagation parameters and conditions in the langasite single crystals has made in the papers [7, 8]. But now the multilayered devices, thin film resonators etc. more often have been used the acoustic waves and substrates with thickness comparable with the wavelength. Also it should point on the trend of including into consideration the new types of acoustic waves, for example, Lamb and SH waves, to extend the acoustic devices possibilities. Devices where such waves are used will be the frequency dispersive ones, i.e. its frequency spectrum depends on the h×f product where h is the plate thickness and f is the frequency, and it was important to take into account the dispersive dependences all of the Lamb and SH wave's parameters. Influence of the dc electric field on propagation of Lamb waves in bismuth germanium oxide piezoelectric thin plates has been analyzed in our preliminary papers [9-11]. The goal of a given paper is to present the results of a complete investigation for zero and first order Lamb and SH waves propagation in a lot of crystalline cuts of langasite plates under the influence of dc electric field.

## THEORY OF LAMB AND SH WAVES PROPAGATION IN PIEZOELECTRIC PLATE UNDER THE INFLUENCE OF HOMOGENEOUS DC ELECTRIC FIELD

Influence of homogeneous dc electric field E on Lamb and SH wave propagation conditions in piezoelectric crystalline plate has been considered on the basis of the theory of bulk acoustic waves propagation in piezoelectric crystals subjected to the action of a bias electric field [5, 6]. Wave equations and electrostatics equation written in the natural state for homogeneously deformed crystals without center of symmetry have the form:

$$\begin{aligned} \rho_0 \ddot{\tilde{U}}_i &= \tilde{\tau}_{ik,k}, \\ \tilde{D}_{m,m} &= 0. \end{aligned} \quad (1)$$

Here $\rho_0$ is the density of crystal taken in non-deformed (initial) state, $\tilde{U}_i$ is the vector of dynamic elastic displacements, $\tau_{ik}$ is the tensor of thermo-dynamical stresses and $D_m$ is the vector of the induction of electricity. Here and further the tilde sign is marked the time dependent variables. Comma after the lower index denotes the spatial derivative. Latin coordinate indexes are changed from 1 to 3. Here and further the summation on twice recurring lower index is understood.

State equations can be written as:

$$\begin{aligned} \tilde{\tau}_{ik} &= C^*_{ikpq} \tilde{\eta}_{pq} - e^*_{nik} \tilde{E}_n, \\ \tilde{D}_n &= e^*_{nik} \tilde{\eta}_{ik} + \varepsilon^*_{nm} \tilde{E}_m, \end{aligned} \quad (2)$$

where $\eta_{AB}$ is the deformation tensor and effective elastic, piezoelectric, dielectric constants are defined by:

$$\begin{aligned} C^*_{iklm} &= C^E_{iklm} + \left( C^E_{iklmpq} d_{jpq} - e_{jiklm} \right) M_j E, \\ e^*_{nik} &= e_{nik} + \left( e_{nikpq} d_{jpq} + H_{njpq} \right) M_j E, \\ \varepsilon^*_{nm} &= \varepsilon^\eta_{nm} + \left( H_{nmik} d_{jik} + \varepsilon^\eta_{nmj} \right) M_j E. \end{aligned} \quad (3)$$

In (3) E is the value of dc electric field applied to the crystal, and $M_j$ is the unit vector of the dc electric field direction; $C^E_{iklmpq}$, $e_{nikpq}$, $\varepsilon^\eta_{nmj}$, $H_{nmik}$ are nonlinear elastic, piezoelectric, dielectric and electrostrictive constants (material tensors) respectively; $d_{jpq}$ and $e_{nik}$ are the piezoelectric tensors, $C_{iklm}^E$ and $\varepsilon_{nm}^\eta$ are elastic and clamped dielectric tensors. Accepting the effective material constants in the form (3) *ipso facto* we have taken into account the so-called physical nonlinearity of the piezoelectric crystalline media. Then substituting (2) into (1) we can obtain Green-Christoffel's equation in a general form which can be used for the analysis of bulk acoustic waves propagation in the case of E-influence.

Let's use the orthogonal coordinate system where the $X_3$ axis directs along the external normal to the surface of a media occupying the space $h \geq X_3 \geq 0$, and the wave propagation direction coincides with $X_1$ axis. Plane waves propagating in the piezoelectric plate are taken in the form:

$$\begin{aligned}\tilde{U}_i &= \alpha_i \exp[i(k_j x_j - \omega t)], \\ \varphi &= \alpha_4 \exp[i(k_j x_j - \omega t)],\end{aligned} \quad (4)$$

where $\alpha_i$ and $\alpha_4$ are amplitudes of elastic wave and electric potential $\varphi$ concerned closely with the wave, and $k_j$ are components of wave propagation vector. Taking into account (2) and (3) the substitution (4) into (1) gives us a specific form equation. So if the electric field is applied to piezoelectric crystal, Green-Christoffel's equation should be written as

$$[\Gamma_{pq}(E) - \rho_0 \omega^2 \delta_{pq}]\tilde{U}_q = 0, \quad (5)$$

where Green-Christoffel's tensor has the form:

$$\begin{aligned}\Gamma_{pq} &= (C_{ipqm}^* + 2d_{jkq}C_{ipkm}^E M_j E)k_i k_m, \\ \Gamma_{q4} &= e_{imq}^* k_i k_m, \\ \Gamma_{4q} &= \Gamma_{q4} + 2e_{ikm}d_{jkq}M_j k_i k_m E, \\ \Gamma_{44} &= -\varepsilon_{nm}^* k_n k_m.\end{aligned} \quad (6)$$

The form (6) of the Green-Christoffel's tensor has been taken into account both the physical and geometrical nonlinearities. The last one is associated with the sample's dimensions and shape changing due to dc electric field influence by means of the direct piezoelectric effect. Propagation of acoustic waves in the piezoelectric plate under the E influence should satisfy to boundary conditions of zero normal components of the stress tensor on the boundaries "crystal-vacuum". Continuity of the electric field components which are tangent to the boundary surface is guaranteed by the condition of the continuity of the electrical potential and normal components of the electric displacement vector:

$$\begin{aligned}\tau_{3k} &= 0, & x_3 &= 0; \; x_3 = h; \\ \varphi &= \varphi^{[I]}, & x_3 &< 0; \\ \varphi &= \varphi^{[II]}, & x_3 &> h; \\ D &= D^{[I]}, & x_3 &< 0; \\ D &= D^{[II]}, & x_3 &> h.\end{aligned} \quad (7)$$

Here the upper index «I» is concerned to the half-space $X_3 > h$ and index «II» – to the half-space $X_3 < 0$. Substituting the solutions (4) into equations (7) and neglecting of the terms which are proportional $E^2$ (and higher order ones), finally we have obtained the system of equations useful to analyze the change of the wave's structure arising as a consequence of crystal symmetry variation and new effective constants appearance:

$$\sum_{n=1}^{8} C_n \left( (C^*_{3ikl} + 2d_{jkp} C^E_{3ipl} M_j E) k^{(n)}_l \alpha^{(n)}_k + e^*_{k3j} k^{(n)}_k \alpha^{(n)}_4 \right) \exp\left(ik^{(n)}_3 h\right) = 0;$$

$$\sum_{n=1}^{8} C_n \left( (e^*_{3kl} + 2d_{jkp} e_{3pl} M_j E) k^{(n)}_l \alpha^{(n)}_k - (\varepsilon^*_{3k} k^{(n)}_k - i\varepsilon_0) \alpha^{(n)}_4 \right) \exp\left(ik^{(n)}_3 h\right) = 0;$$

$$\sum_{n=1}^{8} C_n \left( (C^*_{3ikl} + 2d_{jkp} C^E_{3ipl} M_j E) k^{(n)}_l \alpha^{(n)}_k + e^*_{k3j} k^{(n)}_k \alpha^{(n)}_4 \right) = 0; \qquad (8)$$

$$\sum_{n=1}^{8} C_n \left( (e^*_{3kl} + 2d_{jkp} e_{3pl} M_j E) k^{(n)}_l \alpha^{(n)}_k - (\varepsilon^*_{3k} k^{(n)}_k + i\varepsilon_0) \alpha^{(n)}_4 \right) = 0.$$

Here the index n = 1,…,4 corresponds to the number of one of the partial waves (4) and $C_n$ are the weight coefficients of the partial waves.

It can remember that the equations (8) were obtained at the assumption of homogeneity of applied dc electric field without taking into account the edge effects. But these equations are taken into consideration all the changes both the crystal density and the shape of crystalline sample arising as a consequence of finite deformation of piezoelectric media under the action of strong dc electric field (geometrical nonlinearity) [6].

**ANISOTROPY OF THE DC ELECTRIC FIELD ON THE CHARACTERISTICS OF ACOUSTIC WAVES IN PIEZOELECTRIC PLATE**

Computer simulation of anisotropy of dc electric field influence on the zero and first Lamb and SH waves parameters of LGS piezoelectric plate has made on the basis of the theory given earlier [9-11] and above mentioned dispersive equations. Calculation of phase velocities $v_i$ and $v_{im}$ for the free and metalized surfaces respectively has been carried out. Then the square of electromechanical coupling coefficient (EMCC) has been obtained in a manner of the SAW parameters calculation [13]:

$$K^2 = 2 \frac{v_i - v_{im}}{v_i}, \qquad (9)$$

controlling coefficients of phase velocities by the action of dc electric field E

$$\alpha_{v_i} = \frac{1}{v_i(0)} \left( \frac{\Delta v_i}{\Delta E} \right)_{\Delta E \to 0} \qquad (10)$$

for a lot of crystalline directions has been carried out by our software. Data on material linear and nonlinear electromechanical properties and its temperature coefficients of LGS crystals were taken from [12, 13].

Wave parameters were calculated for the (100), (010), (001) (X, Y, and Z cuts respectively) and (110) crystalline cuts. It was supposed that dc electric field was applied along $X_1$, $X_2$ and $X_3$ axes of special Cartesian coordinate system in which the $X_1$ axis coincides with wave's propagation direction $\vec{N}$ and the $X_3$ axis is parallel to the unit vector $\vec{n}$ normal to free plate surface. All parameters are investigated for free plate surface except the E application along the $X_3$ axes. In the last case it was proposed that the appropriate surface was metalized.

DC electric field influence along the $X_1$ axis, i.e. along the twofold symmetry axis, tends to the decreasing of crystalline symmetry toward monoclinic one (point group of symmetry 2). As a consequence the some new effective constants are induced, but mainly there is the modification of the existing constants:

$$
\begin{aligned}
&C_{11}^* = C_{11} + (C_{111}d_{11} - C_{112}d_{11} + C_{114}d_{11} - e_{111})E; \\
&C_{22}^* = C_{11} + (C_{111}d_{11} + C_{112}d_{11} - 2C_{222}d_{11} - C_{114}d_{14} - 2C_{124}d_{14} - e_{122})E; \quad C_{33}^* = C_{33}; \\
&C_{44}^* = C_{44} + (C_{144}d_{11} - C_{155}d_{11} + C_{444}d_{14} - e_{144})E; \\
&C_{55}^* = C_{44} - (C_{144}d_{11} - C_{155}d_{11} + C_{444}d_{14} - e_{144})E; \\
&C_{66}^* = C_{66} + (-C_{111}d_{11} + C_{222}d_{11} + C_{124}d_{14} - e_{166})E; \\
&C_{12}^* = C_{12} + (-C_{111}d_{11} + C_{222}d_{11} + C_{124}d_{14} - e_{112})E; \quad C_{15}^* = C_{16}^* = C_{25}^* = C_{26}^* = 0; \\
&C_{13}^* = C_{13} + (C_{113}d_{11} - C_{123}d_{11} + C_{134}d_{14} - e_{113})E; \quad C_{35}^* = C_{36}^* = C_{45}^* = C_{46}^* = 0; \\
&C_{14}^* = C_{14} + (C_{114}d_{11} - C_{124}d_{11} + C_{144}d_{14} - e_{114})E; \\
&C_{23}^* = C_{13} + (C_{123}d_{11} - C_{113}d_{11} - C_{134}d_{14} + e_{113})E; \\
&C_{24}^* = -C_{14} + (C_{114}d_{11} + 3C_{124}d_{11} + C_{155}d_{14} - e_{124})E; \quad C_{34}^* = (2C_{134}d_{11} + C_{344}d_{14} - e_{134})E; \\
&C_{56}^* = C_{14} + (2C_{124}d_{11} - 0{,}5C_{144}d_{14} + 0{,}5C_{155}d_{14} - e_{156})E; \quad (11)
\end{aligned}
$$

$$
\begin{aligned}
&e_{11}^* = e_{11} + (e_{111}d_{11} - e_{112}d_{11} + e_{114}d_{14} + H_{11})E; \quad e_{12}^* = -e_{11} + (e_{112}d_{11} - e_{122}d_{11} + e_{124}d_{14} + H_{12})E; \\
&e_{13}^* = (2e_{113}d_{11} + e_{134}d_{14} + H_{13})E; \quad e_{14}^* = e_{14} + (e_{114}d_{11} + e_{124}d_{12} + e_{144}d_{14} + H_{14})E; \\
&e_{25}^* = -e_{14} + (e_{114}d_{11} - e_{124}d_{11} + e_{144}d_{14} + H_{14})E; \quad e_{26}^* = -e_{11} + (e_{111}d_{11} + e_{122}d_{11} - e_{156}d_{14} + H_{66})E \\
&e_{35}^* = (2e_{315}d_{11} + H_{44})E; \quad e_{36}^* = (e_{346}d_{14} + H_{41})E; \\
&e_{15}^* = e_{16}^* = e_{21}^* = e_{22}^* = e_{23}^* = e_{24}^* = e_{31}^* = e_{32}^* = e_{33}^* = e_{34}^* = 0;
\end{aligned}
$$

$$
\begin{aligned}
&\varepsilon_{11}^* = \varepsilon_{11}^\eta + (H_{11}d_{11} - H_{12}d_{11} + H_{14}d_{14} + \varepsilon_{111}^\eta)E; \quad \varepsilon_{22}^* = \varepsilon_{11}^\eta + (H_{12}d_{11} - H_{11}d_{11} - H_{14}d_{14} - \varepsilon_{111}^\eta)E; \\
&\varepsilon_{33}^* = \varepsilon_{33}^\eta; \quad \varepsilon_{23}^* = (2H_{41}d_{11} + H_{44}d_{14})E; \quad \varepsilon_{12}^* = \varepsilon_{13}^* = 0.
\end{aligned}
$$

As it can see of (11) all the effective material constants correspond to the point symmetry group 2. It should be noted that E application leads to the finite static deformation of the crystal because the direct piezoelectric effect.

Phase velocities of the zero and first order Lamb and SH modes propagating along the directions of X-cut are shown on fig. 1 when h×f product is changed from 500 up to 2500 m/s. First order Lamb mode appears if h×f > 1500 m/s. It should be noted that the velocity value of the $A_0$ antisymmetrical mode of Lamb wave is considerably increased by the h×f increasing but simultaneously the EMCC is decreased (fig. 2a). The phase velocity of $S_0$ symmetrical mode is decreased by h×f increasing. There is nonmonotonic EMCC dependence of the mode on h×f product: when $\varphi = 66°$, the EMCC increased up to maximal value $K^2 = 1.03$ % by h×f increasing up to 1500 m/s and then it is decreased by the h×f increasing (fig. 2b). The phase velocity of $SH_0$ mode for the given h×f values is insignificantly decreased by the h×f increasing and EMCC behavior is the same as $S_0$ mode (fig. 2c).

A distinctive feature of plate wave's propagation is the hybridization effect, i.e. the existence of coupled acoustic modes with the energy exchange in the some areas of h×f values [15]. Analysis of zero order waves' propagation in the crystalline plates of the several orientation indicated that as a rule such waves were separately propagated. But the hybridization of zero order waves for a number of crystalline cuts is possible even in the absence of dc electric field. For example the hybridization effect between the $A_0$ and $SH_0$ modes is displayed only in the thick plates when h×f > 2000 m/s in the angles sector from 27° up to 57° (fig. 1d, e). Passing throw the wave hybridization area is accompanied by the sharp variation of phase velocity. More typical situation is one when the hybridization effect between zero and first modes is observed. In particular the hybridization between $S_0$ and $SH_1$ modes is displayed for the rather thick plates if h×f > 1500 m/s (fig. 1c-e). DC electric field application along the $X_1$ or $X_2$ axis intensifies the hybridization effect and as a result the $\alpha_v$ coefficients are exponentially increased and tend to maximal values (fig. 2d-g). DC electric field application along the $X_3$ axis calculated for the both metalized surfaces of plate causes a considerable decreasing of the hybridization effect. When h×f = 500…1500 m/s and if taking close to direction $\varphi = 65°$ the phase velocities of $S_0$ and $SH_0$ modes approach each other and E application intensifies its interaction (fig. 2k, l). But when h×f ≥ 2000 m/s there are observed the anomalously large $\alpha_v$ coefficients with opposite signs in comparison with more thin plates. For example, at this point, when h×f = 500 m/s phase velocity of the $SH_0$ mode is equal to 3408.5 m/s and as a result of E influence its value becomes equal to 3422.8 m/s, i.e. $\alpha_v$ coefficient has a positive sign. When the h×f>1500 m/s phase velocity of the $SH_0$ mode begin to decrease significantly. For example, when h × f = 2500 m/s phase velocity is equal to 2676.0 m/s and as a result of E influence its value becomes equal to 2651.2 m/s, i.e. $\alpha_v$ factor will have a negative sign. The phase velocities of the $S_0$ mode values are close to the QFS phase velocity and $\alpha_v$ coefficient has positive sign. Thus, in the thick plates in the neighborhood of the direction $\varphi = 65°$ the behavior of the zero modes is qualitatively dissimilar in comparison with the thin plates. This can be explained by the convergence of the phase velocities of all the zero Lamb modes and the $SH_1$ mode, mutual exchange of energy between them and the effect of implicit hybridization.

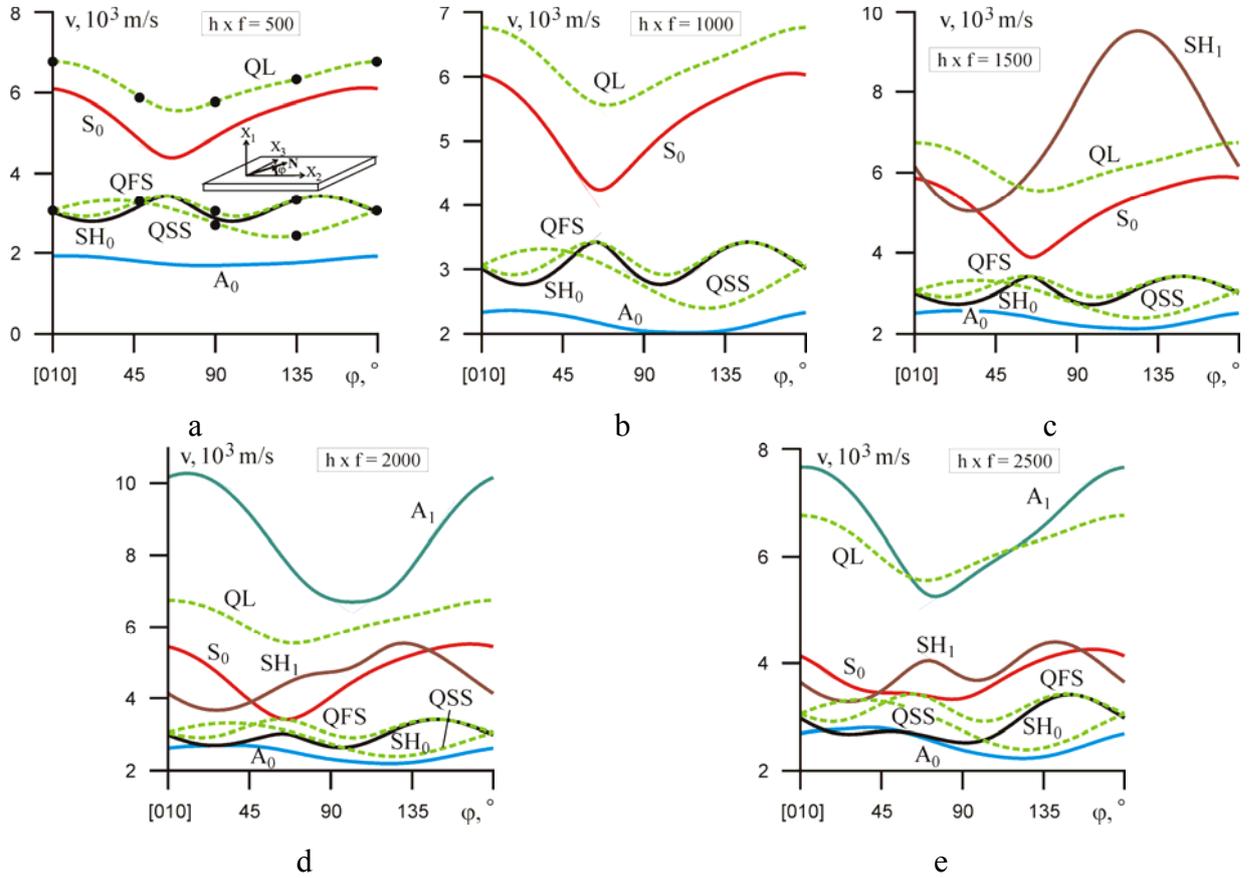

Fig. 1. Phase velocities of acoustical waves propagated in (100) langasite crystalline plate in undisturbed state (E = 0) for a number of h×f values (m/s). Quantities QL, QFS, QSS are designated the curves of quasi-longitudinal, quasi-shear fast and quasi-shear slow bulk acoustic waves respectively. Dots are associated with experimental data [6].

Under the E application along the wave propagation direction (E || $X_1$) the maximal value of $\alpha_v$ coefficient for the $A_0$ mode reaches in the direction close to Y axis: $\alpha_v = -5.9 \cdot 10^{-12}$ m/V (h×f = 2000 m/s) (fig. 2d). The $\alpha_v$ coefficients of the $S_0$ and $SH_0$ modes outside the hybridization areas have the considerably fewer values, so the maximal values are $\alpha_v = -1.74 \cdot 10^{-12}$ m/V (h×f = 1000 m/s) and $\alpha_v = 3.8 \cdot 10^{-12}$ m/V (h×f = 2000 m/s) (fig. 2e, f). If dc electric field is applied along the $X_2$ axis the $\alpha_v$ coefficients of all the modes are fewer by an order of magnitude than in the case E || $X_1$. The E application along the $X_3$ axis, i.e. along the twofold symmetry axis tends to the crystalline symmetry decreasing up to monoclinic one and as a result the $\alpha_v$ coefficients are considerably increased: the maximal values are $\alpha_v = 1.5 \cdot 10^{-10}$ m/V for $S_0$ mode (h×f = 1500 m/s) and $\alpha_v = -1.78 \cdot 10^{-10}$ m/V for $SH_0$ mode (h×f = 2500 m/s).

It should be noted that the anisotropy of the $\alpha_v$ coefficients for first mode $A_1$ (fig. 3a-c) is the similar to ones for the $A_0$ mode (fig. 2d, g, j). But the higher modes possess such characteristic feature as an extreme behavior of $\alpha_v$ coefficients nearby the area where $h \times f$ is close to critical value, and it is possible to observe the appearance of acoustical mode belonging to the higher order than zero one. In this case a small variation of crystalline symmetry and plate configuration tends to the considerable changing of the phase velocity.

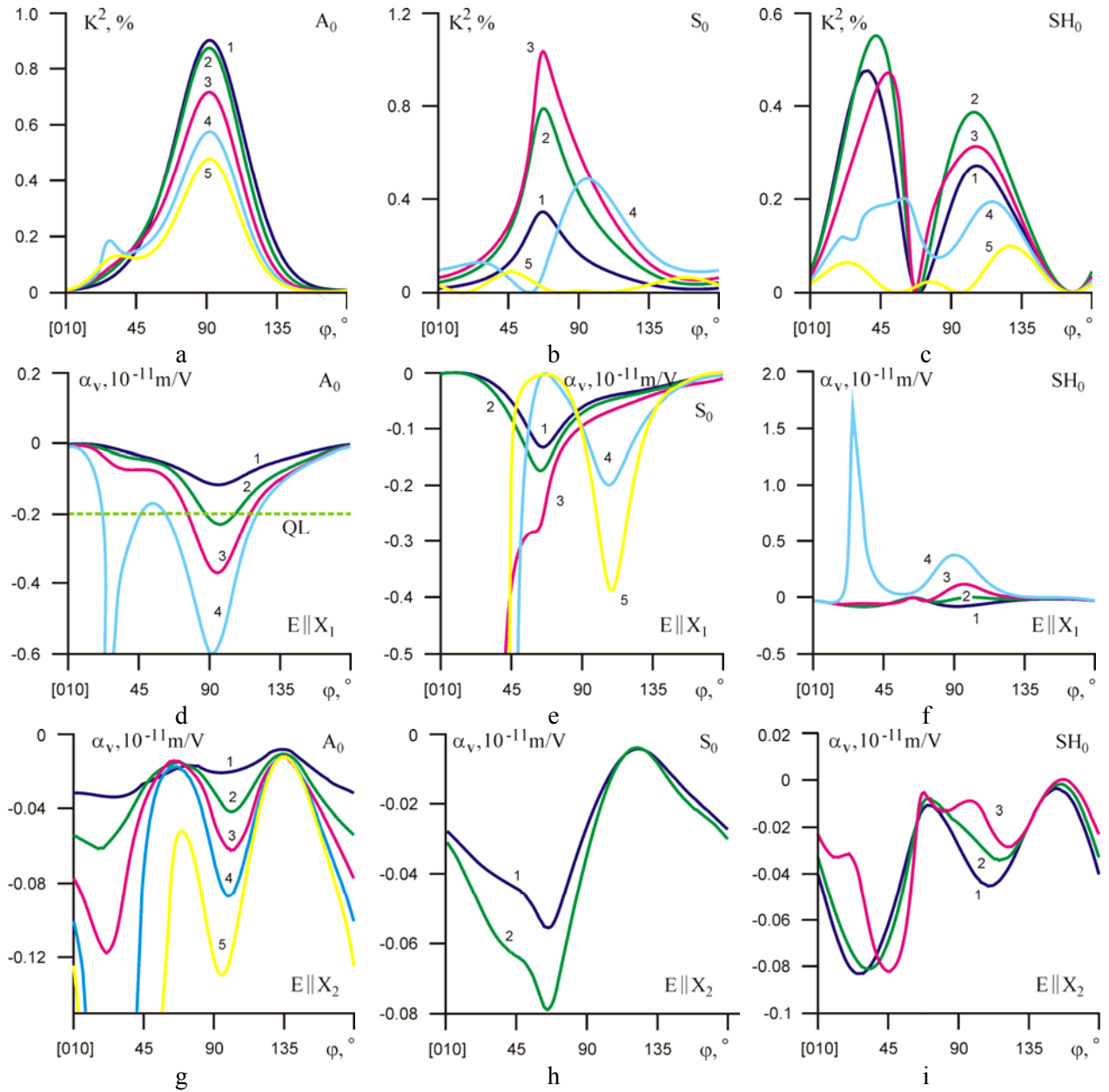

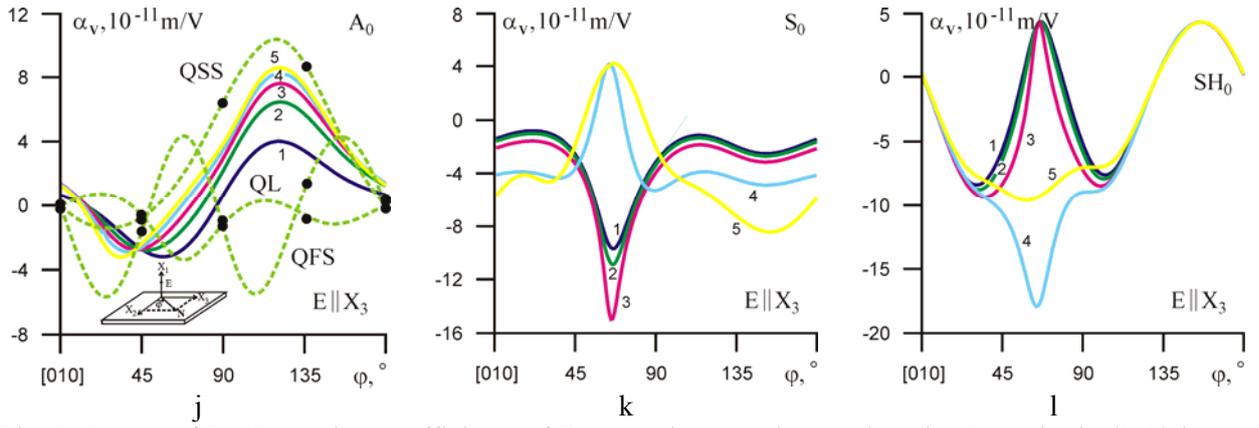

Fig. 2. Square of EMCC and $\alpha_v$ coefficients of BAW and zero order Lamb and $SH_0$ modes in (100) langasite crystalline plane. Numerals are indicated the h×f values (m/s): 1 – h×f = 500; 2 – h×f = 1000; 3 – h×f = 1500; 4 – h×f = 2000; 5 – h×f = 2500. Dots are associated with experimental data [6].

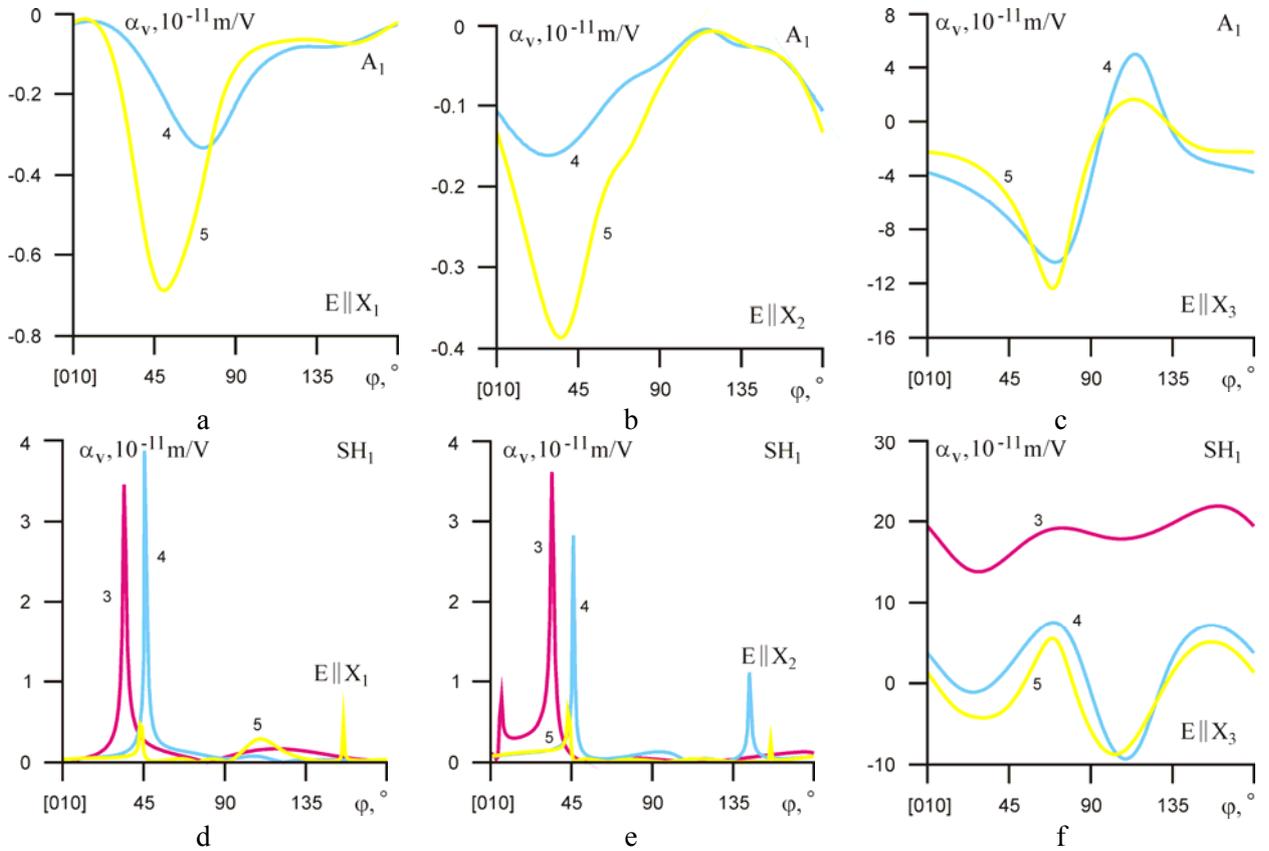

Fig. 3. The $\alpha_v$ coefficients of first order Lamb and $SH_1$ modes in (100) langasite crystalline plane. Numerals are the same as on fig. 2.

In particular taking into account the $SH_1$ mode in the direction $\varphi = 160°$ there is $\alpha_v = 2.19 \cdot 10^{-10}$ m/V if $E \parallel X_3$ (fig. 3f). In addition the hybridization effect tends to exponential dependence of the $\alpha_v$ coefficient for $SH_1$ mode within interaction areas between $SH_1$ and $S_0$ modes shown on fig. 3d, e.

Phase velocities of zero and first order Lamb modes propagating in the Y-cut plate when h×f = 500…3000 m/s are shown on fig. 4. The higher order's modes appear at h×f ≥ 1500 m/s. Phase velocity of the $A_0$ mode is increased by the h×f increasing, but the EMCC is initially in-creased from 0.52 % up to 0.77 % ($\varphi = 0°$), then it is decreased up to 0.32 % in the case $h \times f = 3000$ m/s (fig. 5a). Maximal value EMCC = 2.0 % for the $S_0$ mode of this plate takes place in the $X_1$ propagation direction ($h \times f = 500$ m/s), then the EMCC is decreased up to 0.08 % by the $h \times f$ increasing (fig. 5b). Phase velocities of the $SH_0$ mode are insignificantly decreased by the $h \times f$ increasing, and the EMCC is decreased too from 0.37 % up to 0.02 % (fig. 5c). Hybridization effect between the $S_0$ and $SH_1$ modes appears if $h \times f \geq 2000$ m/s, and the hybridization areas are marked by dashed vertical lines (fig. 4d-f).

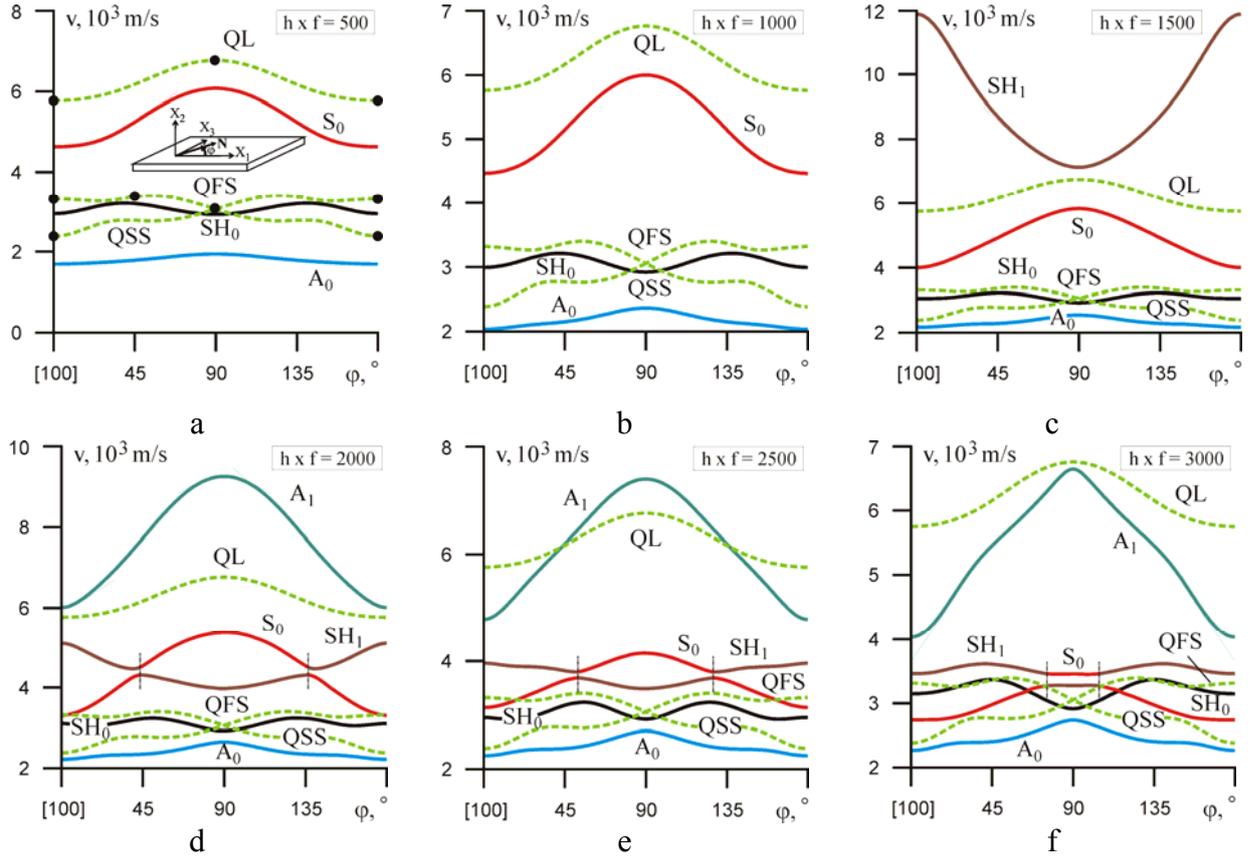

Fig. 4. Phase velocities of acoustical waves propagated in (010) langasite crystalline plate in undisturbed state (E = 0) for a number of $h \times f$ values (m/s). Dots are associated with experimental data [6].

Maximal $\alpha_v$ coefficients of the $A_0$ mode take place if wave propagation direction is close to the [100], and dc electric field coincides with propagation direction (E||$0X_1$), in particular $\alpha_v = 8.3 \cdot 10^{-11}$ m/V if $h \times f = 3000$ m/s. The values of the $\alpha_v$ coefficients for the $S_0$ and $SH_0$ modes outside the hybridization areas are approximately the same as for the $A_0$ mode, and maximal values are $\alpha_v = -1.28 \cdot 10^{-10}$ m/V ($h \times f = 1500$ m/s) and $\alpha_v = -4.11 \cdot 10^{-11}$ m/V ($h \times f = 2000$ m/s) respectively. If dc electric field is applied along the $X_2$ axis the $\alpha_v$ coefficients of the $S_0$ mode have the values fewer on the order of magnitude than in the case E||$X_1$, and the $\alpha_v$ coefficients of the $A_0$ and $SH_0$ modes have the same order of magnitude. The application of dc electric field along the $X_3$ axis increases $\alpha_v$ values significantly in comparison with the case E||$X_2$. Maximal values

$\alpha_v = -8.1 \cdot 10^{-11}$ m/V of the $A_0$ mode take place when $h \times f = 2500$ m/s (fig. 5j). Note that $\alpha_v$ coefficients anisotropy for the $A_0$ and $S_0$ modes is the same as in the case $E \| X_3$.

Qualitative behavior of $\alpha_v$ coefficients anisotropy for the first order plate modes is the similar to one for zero order modes (fig. 6). Peaks on the curves could be explained by the hybridization effect between the $S_0$ and $SH_1$ modes (fig. 6d-f). Behavior of $A_1$ mode when $E \| X_2$ can be explained by the interaction with the $SH_2$ mode (not shown on fig. 4f).

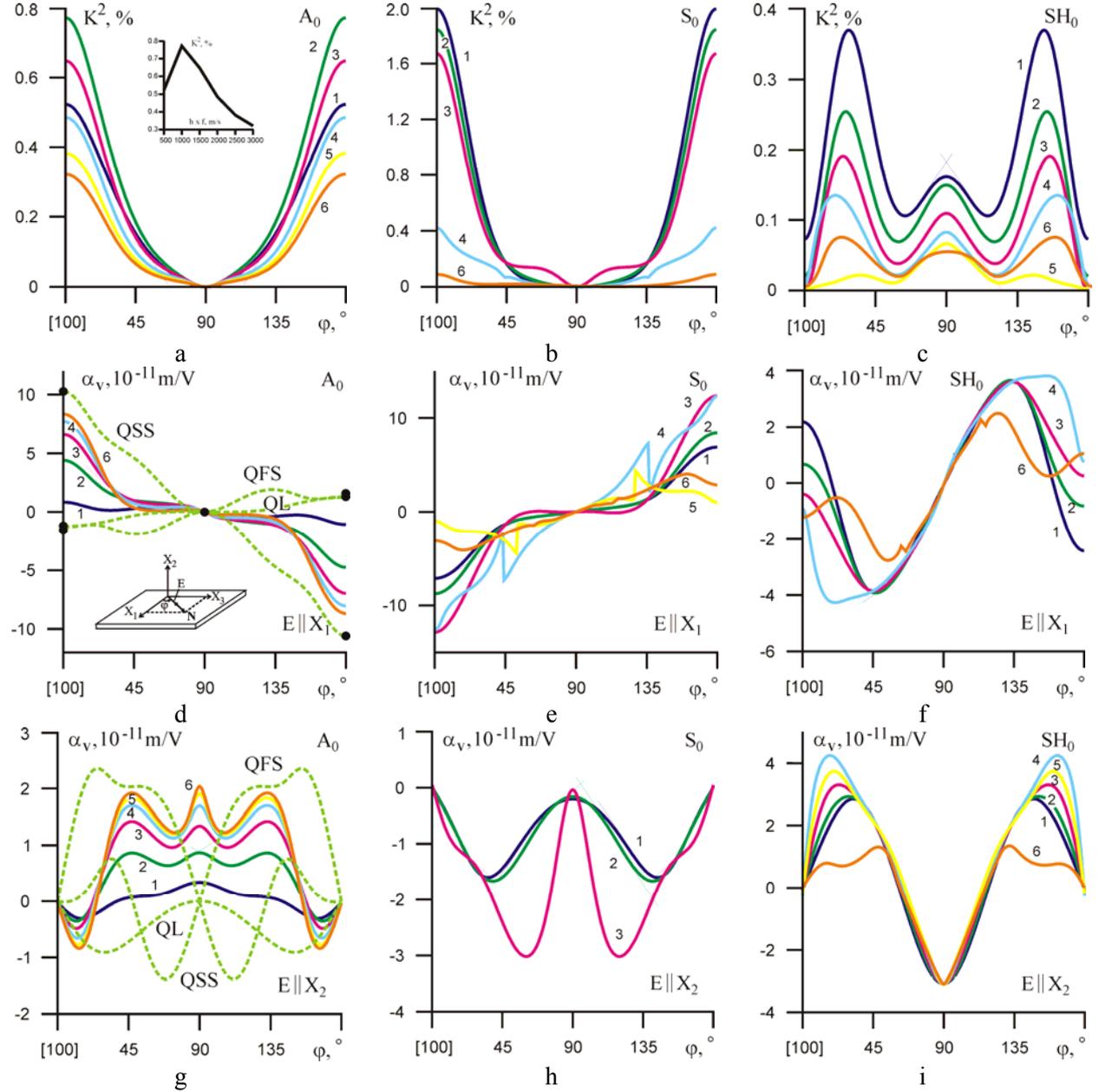

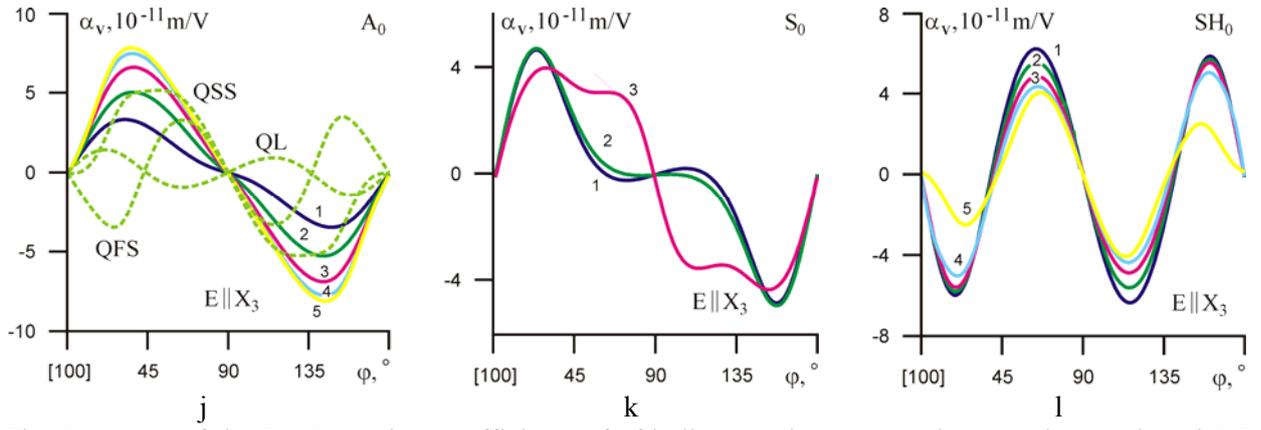

Fig. 5. Square of the EMCC and $\alpha_v$ coefficients of of bulk acoustic waves and zero order Lamb and $SH_0$ modes in (010) langasite crystalline plane. Numerals are indicated the h×f values (m/s): 1 – h×f = 500; 2 – h×f = 1000; 3 – h×f = 1500; 4 – h×f = 2000; 5 – h×f = 2500; 6 – h×f = 3000. Dots are associated with experimental data [6].

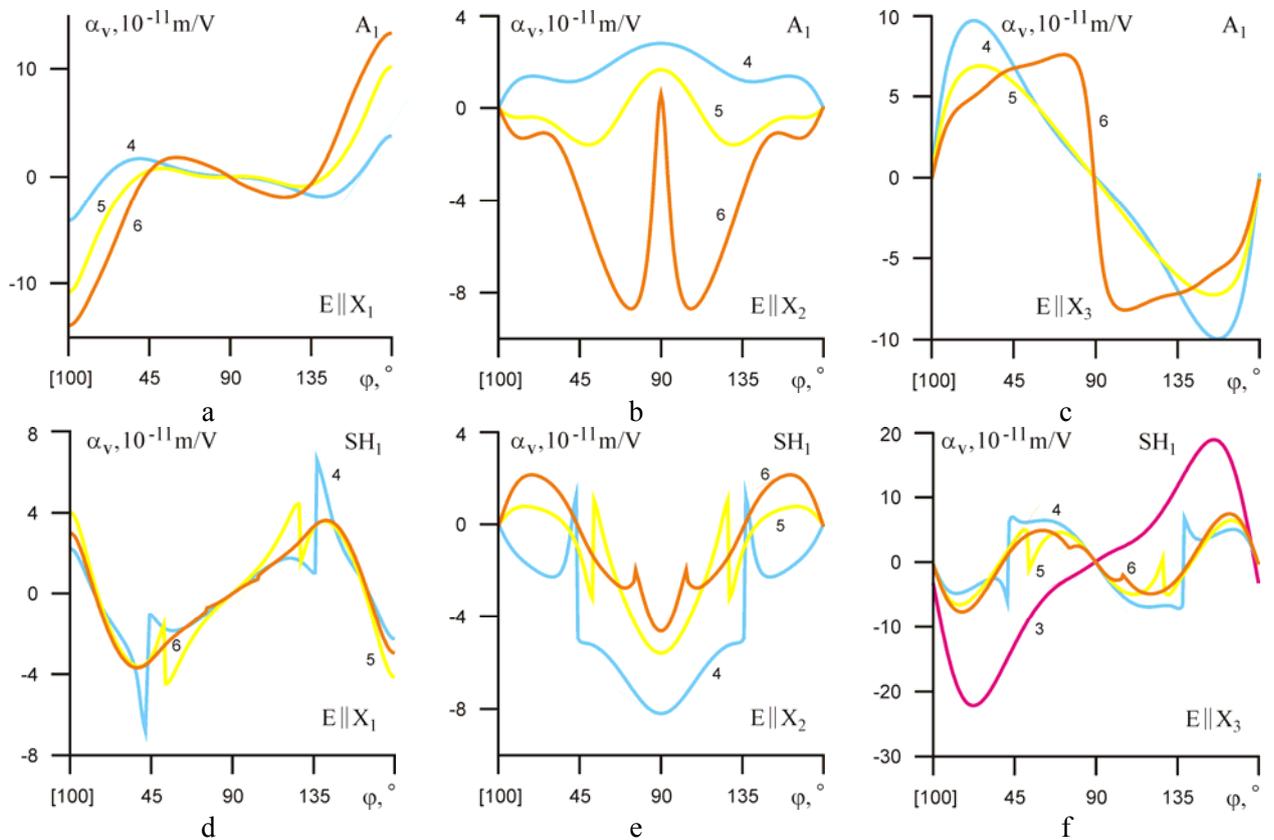

Fig. 6. The $\alpha_v$ coefficients of first order Lamb and $SH_1$ modes in (010) langasite crystalline plane. Numerals are the same as on fig. 5.

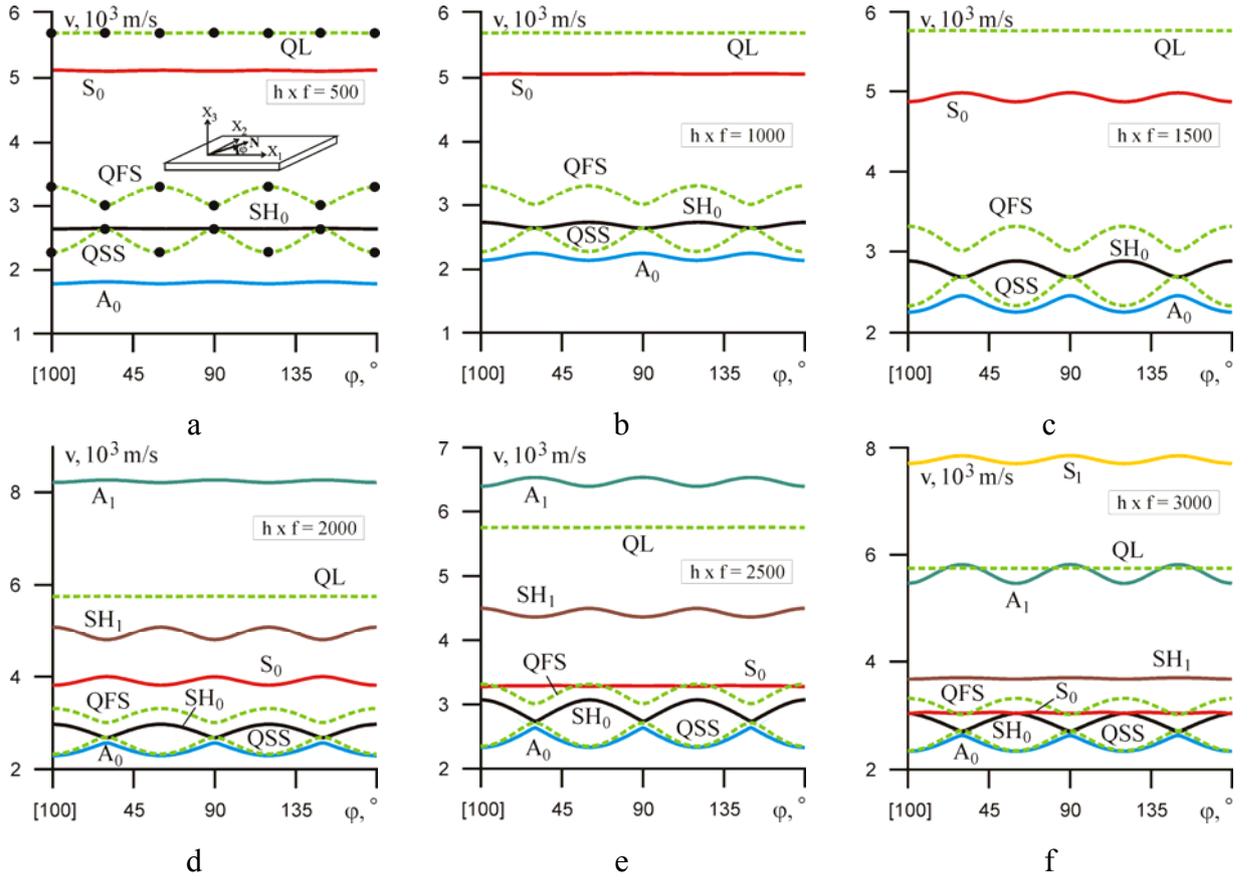

Fig. 7. Phase velocities of acoustical waves propagated in (001) langasite crystalline plate in undisturbed state (E = 0) for a number of h×f values (m/s). Dots are associated with experimental data [6].

Fig. 7 shows the phase velocities of zero and first modes of Lamb waves in a Z-cut at the values h×f = 500…3000 m/s. First and higher order modes appear when h×f≥2000 m/s, and in contrast to the earlier investigated cuts the $S_1$ mode appears only when the value h×f≥3000 m/s. Feature of LGS Z-cut is the absence of an explicit effect of hybridization between any of the modes, since there are no points of the phase velocities equality.

Square of EMCC for the $A_0$ mode increases from 0.03 % (h×f = 500 m/s) to 0.75 % (h×f = 500 m/s) (fig. 8a). With the increasing of h×f values for $SH_0$ and $S_0$ modes the EMCC decreases: for example, from 0.69 % to 0.24 % for $S_0$ mode (fig. 8b). For the $SH_0$ mode in Z-cut the maximum value of the square of EMCC = 2.48 % is achieved in the direction φ = 30° when h×f = 500 m/s, then the EMCC decreases to 0.24 % with the increasing of h×f values (fig. 8c).

It can be noted that for all zero modes in this cut that the behavior of the EMCC and $α_v$ coefficients has been changed qualitatively if h×f ≥ 2000 m/s (fig. 8). This is a consequence of an implicit hybridization effect and the resonant interaction of modes in thick plates when the phase velocities approach each other.

Application of dc electric field along the $X_3$ axis, i.e. in this case threefold symmetry axis leads to a decreasing of symmetry to the point symmetry group 3. However, in this case, the decreasing of symmetry occurs only due to the nonlinear piezoelectric effect and electrostriction (2), i.e. the effect of elastic nonlinearity and static deformation of the crystal are absent. And so new material constants take the following form

$$C^*_{15} = -e_{315}E; \quad C^*_{25} = -e_{325}E; \quad C^*_{46} = -e_{346}E;$$
$$e^*_{16} = H_{41}E; \quad e^*_{15} = H_{44}E; \quad e^*_{21} = H_{41}E; \quad e^*_{22} = H_{42}E; \quad e^*_{24} = H_{44}E; \quad e^*_{31} = H_{31}E. \tag{12}$$

Feature of Z-cut is also the fact that the application of dc electric field along the $X_1$ or $X_2$ axis causes a bigger changes than the E application along the $X_3$ axis which in this case coincides with the $Z\|[001]$ axis (fig. 8d-l).

The behavior of $\alpha_v$ controlling coefficients for the first plate modes is the similar to the zero modes (fig. 9). The changing $\alpha_v$ coefficients for the $SH_1$ mode when $h \times f = 3000$ m/s and $E\|X_2$ can be explained due to the convergence of values of the phase velocities and interaction between the $SH_1$ and $S_0$ modes. Behavior of the $S_1$ mode, which appears only if $h \times f \geq 3000$ m/s, corresponds to the behavior of $S_0$ mode for thin plates.

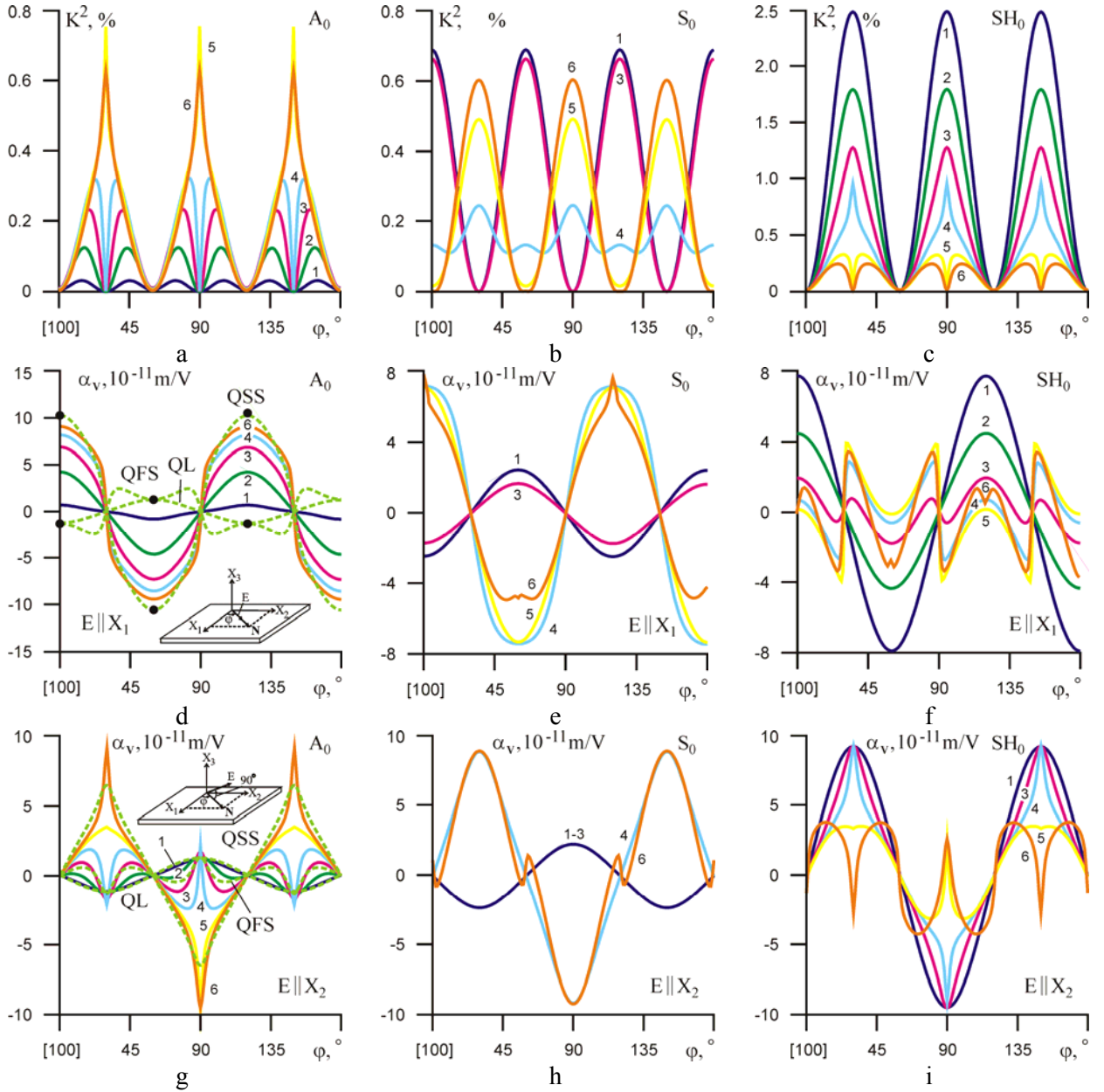

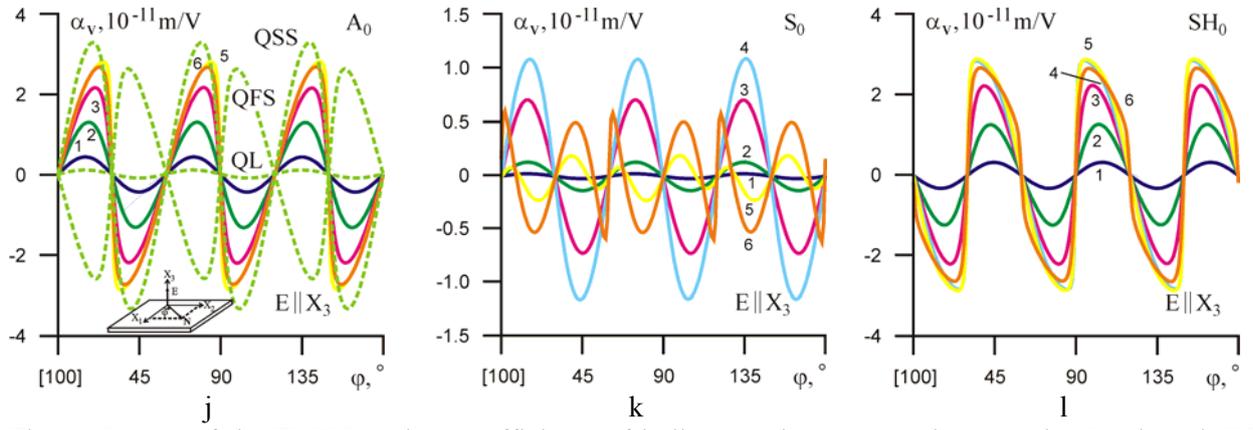

Fig. 8. Square of the EMCC and $\alpha_v$ coefficients of bulk acoustic waves and zero order Lamb and $SH_0$ modes in (001) langasite crystalline plane. Numerals are indicated the h×f values (m/s): 1 – h×f = 500; 2 – h×f = 1000; 3 – h×f = 1500; 4 – h×f = 2000; 5 – h×f = 2500; 6 – h×f = 3000. Dots are associated with experimental data [6].

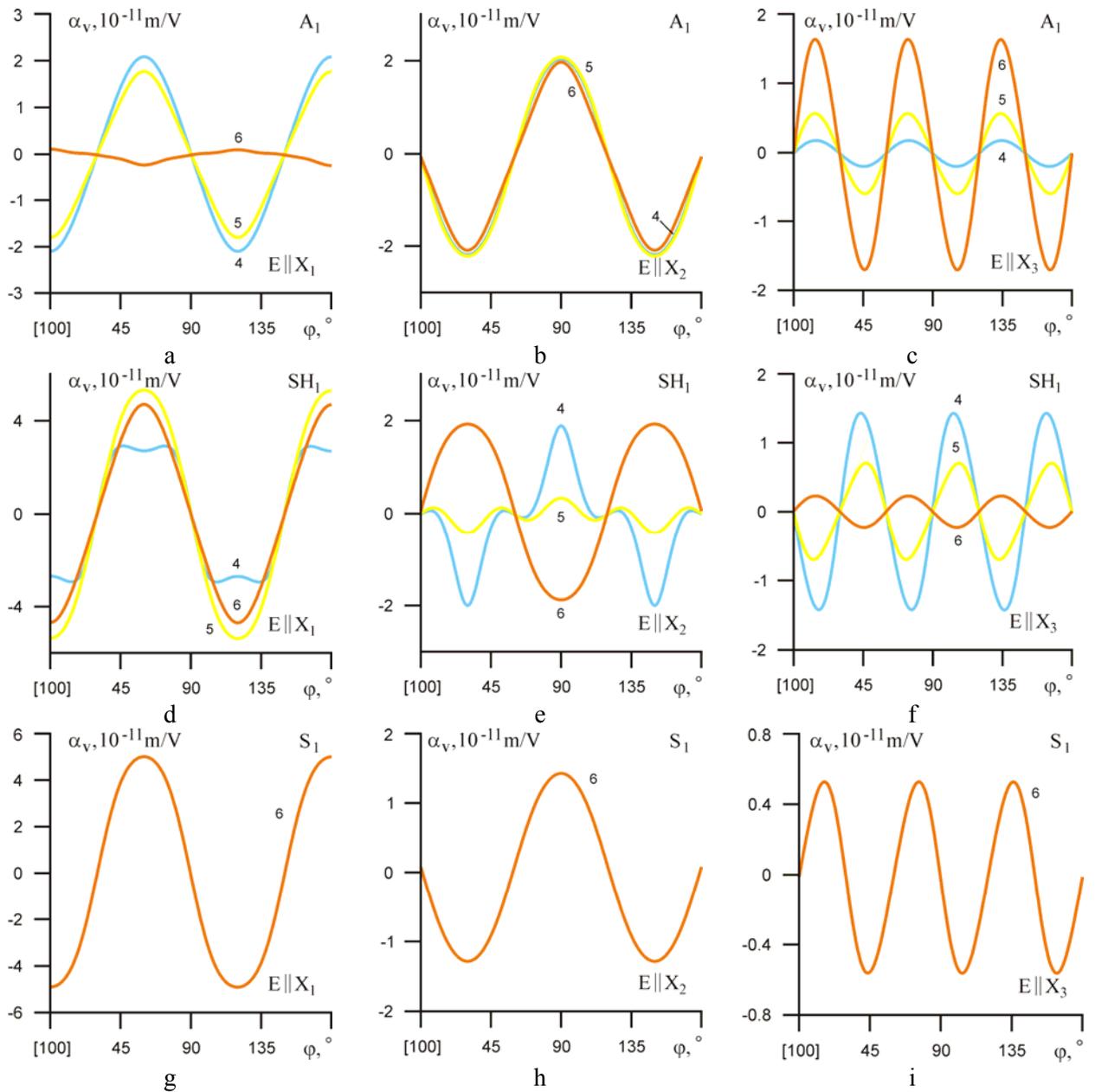

Fig. 9. The $\alpha_v$ coefficients of first order Lamb and $SH_1$ modes in (001) langasite crystalline plane. Numer-

als are the same as on fig. 8.

Fig. 10 shows the phase velocities of zero and first modes of Lamb waves in Y+45° cut plate when h×f = 500…2500 m/s. The effect of hybridization between the $S_0$ and $SH_1$ modes takes place for values h×f >1500 m/s, and areas of hybridization are marked by vertical dashed lines (fig. 10d-e). With the increasing of h×f values the EMCC for the $A_0$ mode is initially increased from 0.7 % to 0.82 % (φ = 0°), then it is decreased to 0.38 % at h×f = 2500 m/s (fig. 11a). It should be noted that the $S_0$ mode has a maximum value of the EMCC square $K^2$ =1.19 % in the direction defined by the angle φ = 11° from the [110] direction in the (110) plane (h×f = 1500 m/s) (fig. 11b). Phase velocities of $SH_0$ mode for selected area of h×f values are slightly changed with the h×f increasing, and the square of the EMCC is decreased (fig. 11c).

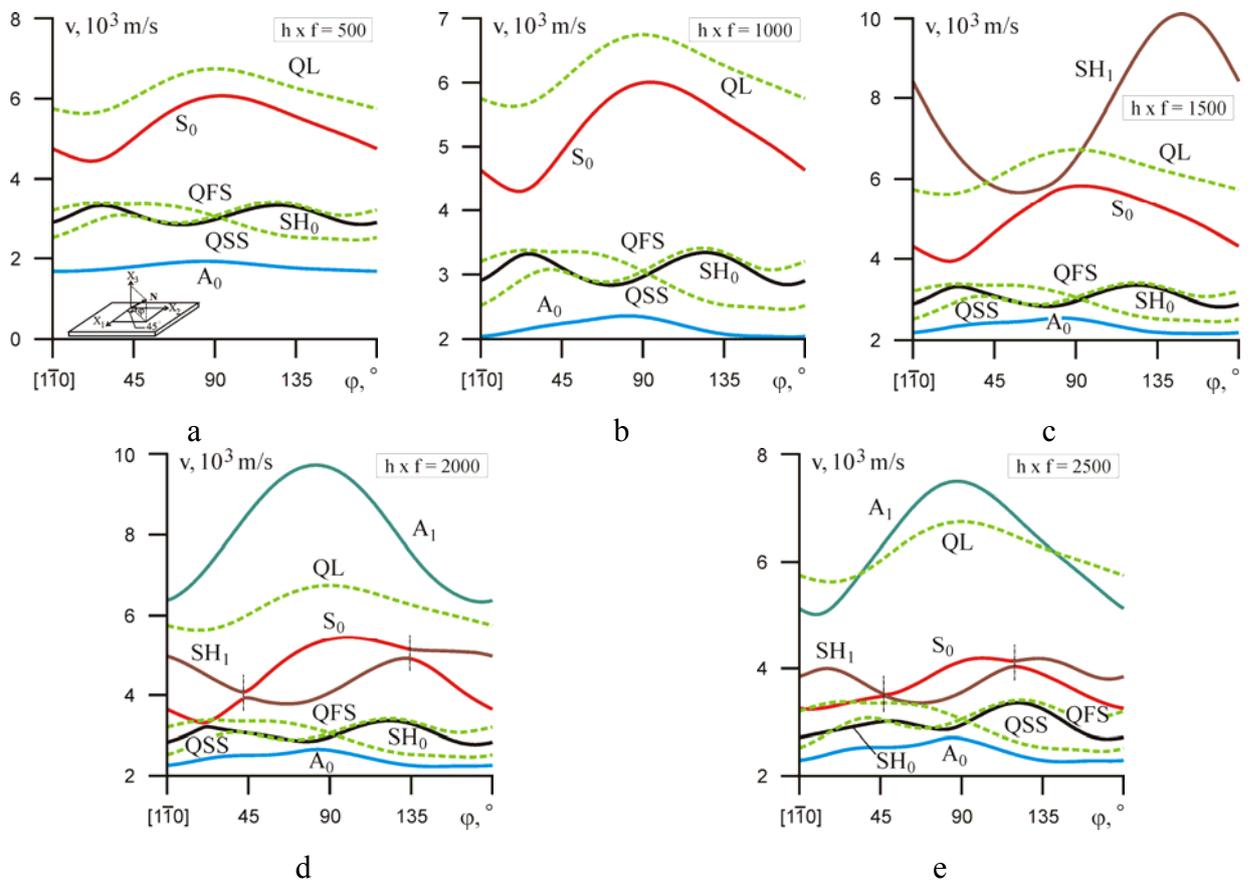

Fig. 10. Phase velocities of acoustical waves propagated in (110) langasite crystalline plate in undisturbed state (E = 0) for a number of h×f values (m/s).

When dc electric field is applied along the direction of wave propagation (E∥$X_1$) maximal $α_v$ values of the $A_0$ mode are achieved in the vicinity of the [110] propagation direction, in particular $α_v$ = -8.0·$10^{-11}$ m/V when h×f = 2500 m/s (fig. 11d). The $α_v$ coefficients of the $S_0$ and $SH_0$ modes outside the hybridization areas have the order of magnitude and the maximal values are $α_v$ = 8.71·$10^{-11}$ m/V (h×f = 1500 m/s) and $α_v$ = -7.37·$10^{-11}$ m/V (h×f = 2500 m/s) respectively. However, the effect of hybridization, as described above, tends to the $α_v$ increasing in hybridization area and changes the EMCC. The $α_v$ anisotropy for the $SH_0$ mode (fig. 11f) when dc electric field is ap-plied along the $X_1$ axis is the similar one as a $S_0$ mode (fig. 11e). When dc electric field is applied along the $X_2$ axis the $\alpha_v$ values for all modes are much smaller than in the case $E \| X_1$.

Application of dc electric field along the $X_3$ axis significantly increases the $\alpha_v$ values in comparison with the case $E \| X_2$. Maximal value $\alpha_v = 1.01 \cdot 10^{-10}$ m/V is achieved for the $S_0$ mode at $h \times f = 1500$ m/s (fig. 11k).

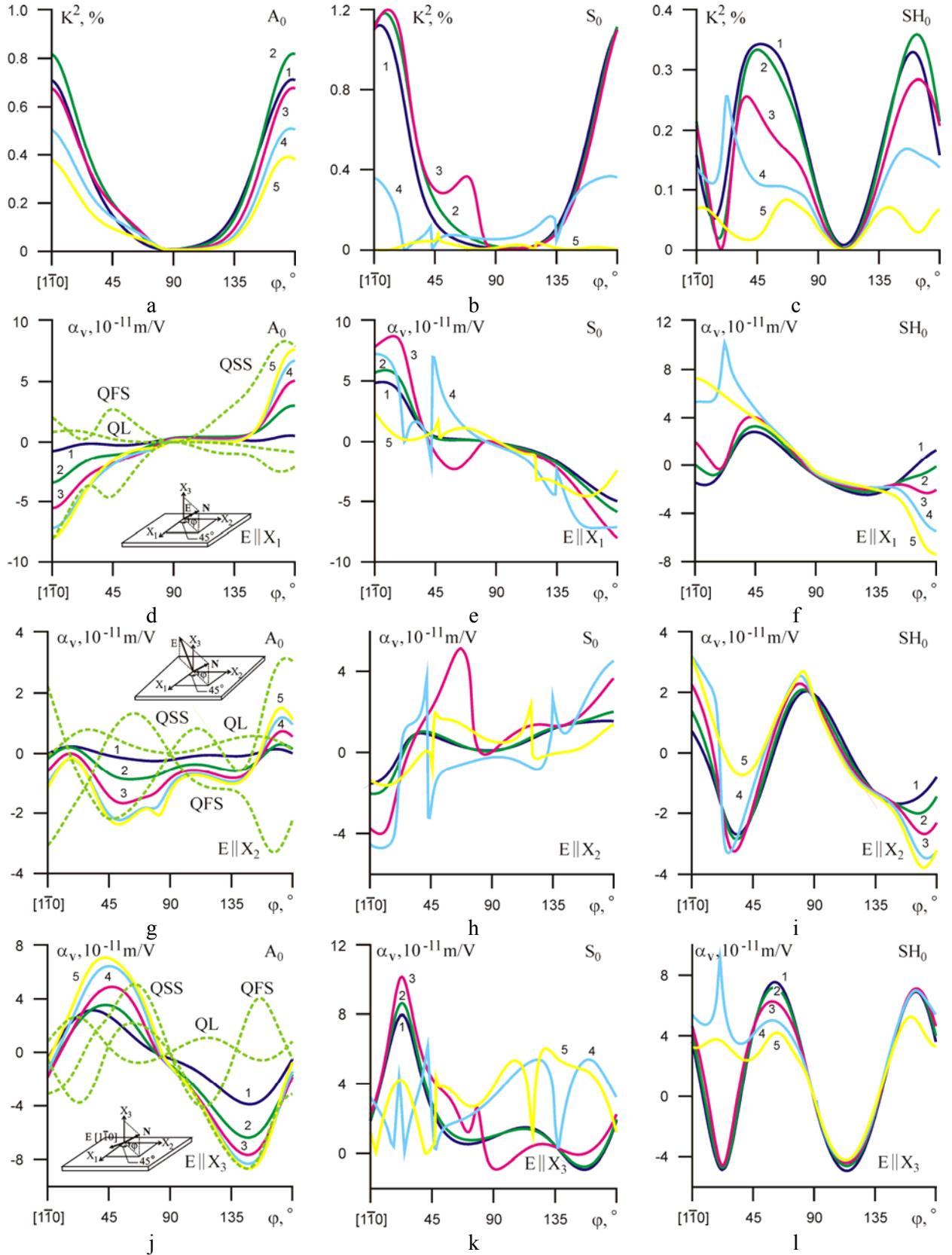

Fig. 11. Square of the EMCC and $\alpha_v$ coefficients of bulk acoustic waves and zero order Lamb and $SH_0$ modes in (110) langasite crystalline plane. Numerals are indicated the h×f values (m/s): 1 – h×f = 500; 2 – h×f = 1000; 3 – h×f = 1500; 4 – h×f = 2000; 5 – h×f = 2500; 6 – h×f = 3000. Dots are associated with experimental data [6].

The $α_v$ coefficients for the first plate modes exceed substantially the ones in the case of zero modes (fig. 12). This is explained by a greater magnitude of the phase velocity of these waves. For example, the $α_v$ values of the $SH_1$ mode at $h×f = 1500$ m/s (moment of Lamb wave appearance when the $h×f$ value is equal to the critical one) for all the dc electric field orientations exceed more than two times the $α_v$ values of the more thick plates in which phase velocity has a significantly smaller value. For plates with $h×f > 2000$ m/s the effect of hybridization between the $S_0$ and $SH_1$ modes significantly affects on the $α_v$ coefficients.

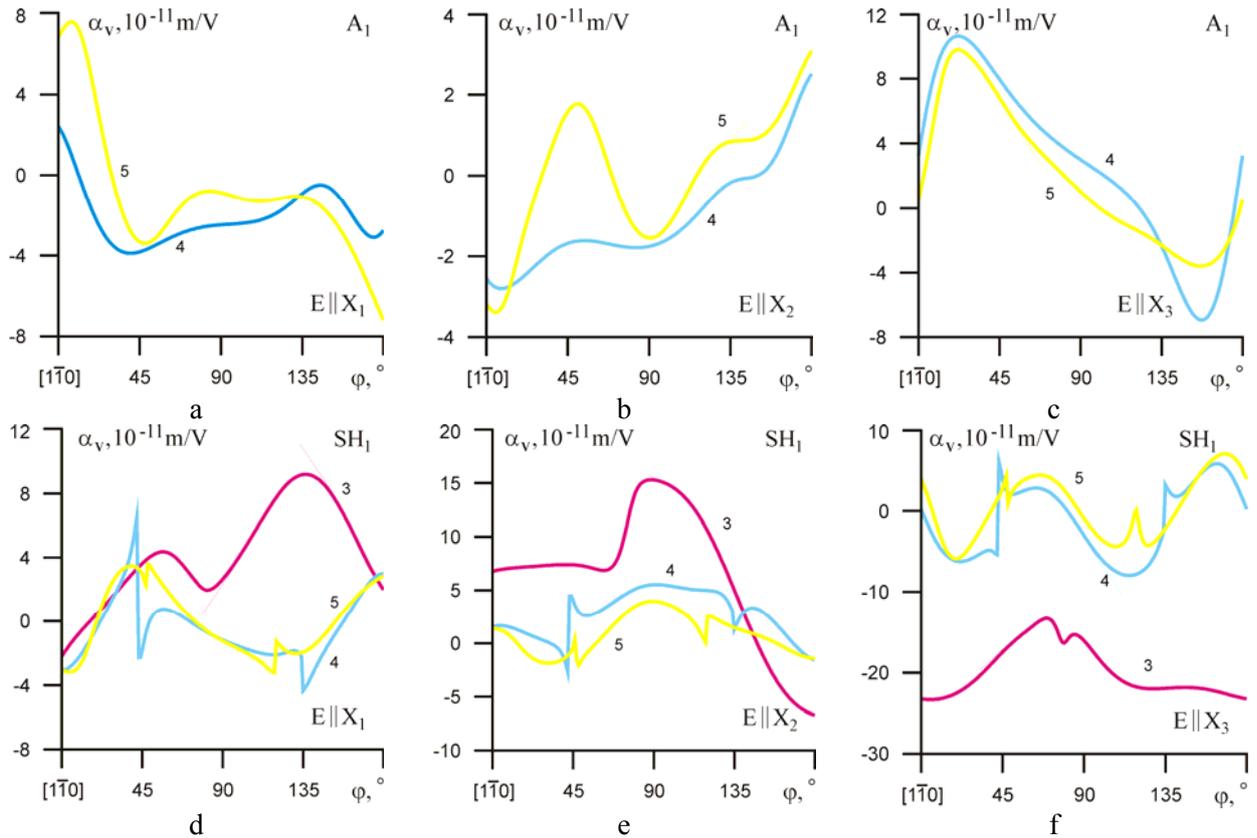

Fig. 12. The $α_v$ coefficients of first order Lamb and $SH_1$ modes in (110) langasite crystalline plane. Numerals are the same as on fig. 11.

Extreme values of the $α_v$ coefficients for the Lamb and SH waves are shown in the table 1.

**Table 1. The maximum and minimum values of the $\alpha_v$ coefficients for the Lamb and SH waves**

| Crystalline plane | Mode | Orientation of the dc electric field | Angle $\varphi$,° | h×f, m/s | $\alpha_v$, $10^{-11}$ m/V |
|---|---|---|---|---|---|
| 1 | 2 | 3 | 4 | 5 | 6 |
| X-cut | $A_0$ | $E\|\|X_3$ | 121 | 2500 | 8.59 |
| | | $E\|\|X_1$ | 9 | 500 | 0 |
| | $SH_0$ | $E\|\|X_3$ | 65 | 2000 | -17.87 |
| | | $E\|\|X_2$ | 159 | 1500 | 0 |
| | $S_0$ | $E\|\|X_3$ | 66 | 1500 | -14.93 |
| | | $E\|\|X_1$ | 9 | 1000 | 0 |
| | $A_1$ | $E\|\|X_3$ | 70 | 2500 | -12.32 |
| | | $E\|\|X_2$ | 119 | 2500 | -0.01 |
| | $SH_1$ | $E\|\|X_3$ | 160 | 1500 | 21.89 |
| | | $E\|\|X_1$ | 75 | 2500 | 0 |
| Y-cut | $A_0$ | $E\|\|X_1$ | 180 | 3000 | -8.69 |
| | | $E\|\|X_1$ | 90 | 500 | 0 |
| | $SH_0$ | $E\|\|X_3$ | 117 | 500 | -6.36 |
| | | $E\|\|X_3$ | 90 | 1500 | 0 |
| | $S_0$ | $E\|\|X_1$ | 0 | 1500 | -12.87 |
| | | $E\|\|X_1$ | 90 | 500 | 0 |
| | $A_1$ | $E\|\|X_1$ | 0 | 3000 | -13.88 |
| | | $E\|\|X_1$ | 90 | 2500 | 0 |
| | $SH_1$ | $E\|\|X_3$ | 23 | 1500 | -22.17 |
| | | $E\|\|X_2$ | 0 | 2000 | 0 |
| Z-cut | $A_0$ | $E\|\|X_2$ | 90 | 3000 | -9.54 |
| | | $E\|\|X_3$ | 60 | 500 | -0.01 |
| | $SH_0$ | $E\|\|X_2$ | 90 | 2000 | -9.56 |
| | | $E\|\|X_1$ | 42 | 1500 | 0 |
| | $S_0$ | $E\|\|X_2$ | 90 | 3000 | -9.24 |
| | | $E\|\|X_3$ | 27 | 500 | 0 |
| | $A_1$ | $E\|\|X_2$ | 150 | 2500 | -2.22 |
| | | $E\|\|X_1$ | 94 | 3000 | 0 |
| | $SH_1$ | $E\|\|X_1$ | 120 | 2500 | -5.38 |
| | | $E\|\|X_3$ | 30 | 3000 | 0 |
| | $S_1$ | $E\|\|X_1$ | 60 | 3000 | 5.02 |
| | | $E\|\|X_2$ | 59 | 3000 | 0 |

| 1 | 2 | 3 | 4 | 5 | 6 |
|---|---|---|---|---|---|
| Y+45°-cut | $A_0$ | $E \| X_3$ | 148 | 2500 | -8.67 |
| | | $E \| X_3$ | 81 | 1500 | -0.02 |
| | $SH_0$ | $E \| X_3$ | 61 | 500 | 7.55 |
| | | $E \| X_2$ | 111 | 1500 | -0.02 |
| | $S_0$ | $E \| X_3$ | 24 | 1500 | 10.07 |
| | | $E \| X_2$ | 90 | 1500 | 0.01 |
| | $A_1$ | $E \| X_3$ | 22 | 2000 | 10.62 |
| | | $E \| X_2$ | 150 | 2000 | 0 |
| | $SH_1$ | $E \| X_3$ | 7 | 1500 | -23.31 |
| | | $E \| X_2$ | 159 | 2500 | -0.02 |

## CONCLUSION

Using previously obtained langasite data on material properties, the analysis of anisotropy characteristics of Lamb and SH waves (phase velocities, electromechanical coupling coefficients, and the controlling coefficients of the phase velocity under the dc electric field influence) in a piezoelectric plate subjected to the action of the homogeneous electric field has been carried out. Nature of dispersion behavior of acoustic modes has been analyzed in detail. Crystalline directions and cuts with maximal and minimal influence of dc electric field have indicated in langasite. It was shown that under the certain orientations of dc electric field and by the increasing of the plate thickness the interaction between some modes may occur. Most promising directions and cuts to be used for the design of acoustoelectronics devices have identified. For example let's take the langasite X-cut plate there are useful parameters of the $S_0$ mode: phase velocity v = 3837.71 m/s in the propagation direction $\varphi = 65°$ at h×f = 1500 m/s; acoustic wave path 0.01 m; delay time $\tau$ = 2.60572 mcsec; $\alpha_v$ = -1.47·$10^{-10}$ m/V. As a result the dc electric field application ($\Delta E = 1 \cdot 10^7$ V/m) along the normal to metalized surface ($E \| X_3$) will cause the delay variation $\Delta t = \pm 3.85$ nsec.

This paper was supported by grant N 4645.2010.2 (Science Schools) of President of Russia.

## REFERENCES

1. I. M. Silvestrova, Yu. V. Pisarevsky, I. A. Senyuschenkov, and A. I. Krupnyi, "Temperature dependences of elastic properties of single crystal $La_3Ga_5SiO_{14}$," *Physics of the Solid State*, vol. 28, no. 9, pp. 2875–2878, 1986.
2. B. V. Mill and Yu. V. Pisarevsky, "Langasite-type materials: from discovery to present state," *Proc. IEEE/EIA Freq. Contr. Symp.*, pp. 133-144, 2000.

3. R. Fachberger, G. Bruckner, G. Knoll, et al, "Applicability of LiNbO$_3$, langasite and GaPO$_4$ in high temperature SAW sensors operating at radio frequencies," *IEEE Trans. Ultrason. Ferroel. Freq. Contr.*, vol. 51, no. 11, pp. 1427-1431, 2004.

4. M. Schulz, D. Richter, and H. Fritze, "Material and resonator design dependant loss in langasite bulk acoustic wave resonators at high temperatures" *Proc. IEEE Interl Ultrason. Symp.*, pp. 1676-1679, 2009.

5. M. P. Zaitseva, Yu. I. Kokorin, Yu. M. Sandler, V. M. Zrazhevsky, B. P. Sorokin, and A. M. Sysoev, "Non-linear electromechanical properties of acentric crystals," Novosibirsk: Nauka, Siberian Branch, 1986, 177 p.

6. K. S. Aleksandrov, B. P. Sorokin, and S. I. Burkov, "Effective piezoelectric crystals for acoustoelectronics, piezotechnics and sensors," vol. 2. Novosibirsk: SB RAS Publishing House, 2008, 429 p.

7. K. S. Aleksandrov, B. P. Sorokin, P. P. Turchin, S. I. Burkov, D. A. Glushkov, and A. A. Karpovich, "Effects of static electric field and of mechanic pressure on surface acoustic waves propagation in La$_3$Ga$_5$SiO$_{14}$ piezoelectric single crystals," *Proc. of 1995 IEEE Ultrason. Symp.*, vol. 1, pp. 409-412, 1995.

8. B. P. Sorokin, P. P. Turchin, S. I. Burkov, D. A. Glushkov, and K. S. Aleksandrov, "Influence of static electric field, mechanic pressure and temperature on the propagation of acoustic waves in La$_3$Ga$_5$SiO$_{14}$ piezoelectric single crystals," *Proc. of 1996 IEEE Intl Freq. Contr. Symp.*, pp. 161-169, 1996.

9. S. I. Burkov, O. P. Zolotova, and B.P. Sorokin, "Influence of the external electric field on propagation of Lamb waves in thin piezoelectric sheets," *Proc. 2008 IEEE Intl Ultrason. Symp.*, pp. 1812-1814, 2008.

10. S. I. Burkov, O. P. Zolotova, B. P. Sorokin, and K. S. Aleksandrov, "Effect of external electrical field on characteristics of a Lamb wave in a piezoelectric plate," *Acoustical Physics*, vol. 56, no. 5, pp. 644–650, 2010.

11. O. P. Zolotova, S. I. Burkov, and B. P. Sorokin, "Propagation of the Lamb and SH-waves in piezoelectric cubic crystal's plate," *J. of Siberian Federal University, Mathematics&Physics*, vol. 3, no. 2, pp. 185-204, 2010.

12. K. S. Aleksandrov, B. P. Sorokin, P. P. Turchin, and D. A. Glushkov, "Non-linear piezoelectricity in La$_3$Ga$_5$SiO$_{14}$ piezoelectric single crystal," *Ferroelectr. Lett.*, vol. 14, no. 5-6, pp. 115–125, 1992.

13. B. P. Sorokin, P. P. Turchin, and D. A. Glushkov, "The elastic nonlinearity and peculiarities of bulk acoustic wave propagation under the influence of homogeneous mechanical stresses in $La_3Ga_5SiO_{14}$ single crystals," *Physics of the Solid State*, vol. 36, no. 10, pp. 2907–2916, 1994.

14. B. D. Zaitsev, I. E. Kuznetsova, I. A. Borodina, and S.G. Joshi, "Characteristics of acoustic plate waves in potassium niobate ($KNbO_3$) single crystal", *Ultrasonics*, vol. 39, pp. 51-55, 2001.

15. I. E. Kuznetsova, B. D. Zaitsev, A. A. Teplykh, and I. A. Borodina, "Hybridization of acoustic waves in piezoelectric plates", *Acoustical Physics*, vol. 53, no. 1, pp. 64-69, 2007.